\documentclass[11pt]{article}
\usepackage{latexsym}
\usepackage{amsmath}

 \csname
@addtoreset\endcsname{equation}{section}

\def\pe{\prime}
\def\3s{{s \choose 3}}
\def\4s{{s \choose 4}}
\def\5s{{s \choose 5}}
\def\6s{{s \choose 6}}
\def\12{\frac{1}{2}}
\def\fr{\frac}
\def\pr{\partial}
\def\prd{\partial \cdot}

\def\bec{\begin{center}}
\def\ec{\end{center}}
\def\a{\alpha}  

\def\b{\beta}

\def\d{\delta} 

\def\e{\epsilon}

\def\vf{\varphi}

\def\l{\lambda}
\def\L{\Lambda}
\def\m{\mu}

\def\r{\rho}

\def\h{\eta}

\def\cB{{\cal B}}
\def\cG{{\cal G}}

\def\cJ{{\cal J}}
\def\cC{{\cal C}}
\def\cL{{\cal L}}

\def\cF{{\cal F}}

\def\cA{{\cal A}}
\def\cW{{\cal W}}
\def\cZ{{\cal Z}}

\def\cR{{\cal R}}
\def\cS{{\cal S}}

\def\cH{{\cal H}}

\def\cA{{\cal A}}

\def\be{\begin{equation}}
\def\ee{\end{equation}}
\def\bea{\begin{eqnarray}}
\def\eea{\end{eqnarray}}
\def\ba{\begin{array}}
\def\ea{\end{array}}

\def\scs#1{\section{\sc #1}}

\def\dsll{\not {\! \pr}}
\def\psisl{\not {\! \! \psi}}

\def\cWsl{\not {\! \! \cal W}}
\def\e{\epsilon}

\def\esl{\not {\! \epsilon}}
\def\ssl{\not {\! \cal S}}
\def\xisl{\not {\! \xi}}
\def\esl{\not {\! \epsilon}}
\def\ssl{\not {\! \cal S}}

\allowdisplaybreaks

\begin{document}
\begin{flushright}
RM3-TH/05-4 \\
ROM2F-05/13\\

\end{flushright}

\vspace{20pt}

\begin{center}


{\Large\sc Minimal Local Lagrangians for Higher-Spin Geometry}\\


\vspace{30pt} {\sc Dario Francia} \\ \vspace{5pt} {\sl\small Dipartimento di Fisica\\ Universit\`a di Roma Tre\\
I.N.F.N., Sezione di Roma Tre \\ Via della Vasca Navale 84, I-00146
Roma, Italy \\ e-mail: {\small \it
francia@fis.uniroma3.it}}\vspace{10pt}

and \vspace{10pt}

{\sc Augusto Sagnotti}\\ \vspace{5pt} {\sl\small Dipartimento di Fisica\\
Universit\`a di Roma ''Tor Vergata''\\ I.N.F.N., Sezione di Roma
''Tor Vergata''
\\ Via della Ricerca Scientifica 1, I-00133 Roma, Italy\\ e-mail: {\small \it
sagnotti@roma2.infn.it} }


\vspace{30pt} {\sc\large Abstract}\end{center}
The Fronsdal Lagrangians for free totally symmetric rank-$s$ tensors
$\vf_{\mu_1 \ldots \; \mu_s}$ rest on suitable trace constraints for
their gauge parameters and gauge fields. Only when these constraints
are removed, however, the resulting equations reflect the expected
free higher-spin geometry. We show that geometric equations, in both
their local and non-local forms, can be simply recovered from local
Lagrangians with \emph{only two} additional fields, a rank-$(s-3)$
compensator $\alpha_{\mu_1 \ldots \; \mu_{s-3}}$ and a rank-$(s-4)$
Lagrange multiplier $\beta_{\mu_1 \ldots \; \mu_{s-4}}$. In a
similar fashion, we show that geometric equations for unconstrained
rank-$n$ totally symmetric spinor-tensors $\psi_{\mu_1 \ldots \;
\mu_n}$ can be simply recovered from local Lagrangians with
\emph{only two} additional spinor-tensors, a rank-$(n-2)$
compensator $\xi_{\mu_1 \ldots \; \mu_{n-2}}$ and a rank-$(n-3)$
Lagrange multiplier $\lambda_{\mu_1 \ldots \; \mu_{n-3}}$.
\vfill
 \setcounter{page}{1}



\newpage


\scs{Introduction and Summary}\label{sec:Intro}


Higher-spin gauge fields\footnote{The web site
http://www.ulb.ac.be/sciences/ptm/pmif/Solvay1proc.pdf contains the
Proceedings of the First Solvay Workshop on Higher-Spin Gauge
Fields, with a number of contributions, including
\cite{bd,fh,bcs,bkiv,sss} that are more closely related to this
work, and many references to the original literature.} are an old
and fascinating corner of Field Theory, with many open questions,
some of which appear of direct relevance for a deeper understanding
of String Theory. Only a few Lorentz representations play a role in
the present models of the Fundamental Interactions (two-tensors,
vectors, scalars, spinors, and often their super-partners), while
the term ``higher spin'' qualifies in principle fields belonging to
all types of representations that are more complicated than these.
In practice, however, one often restricts the attention to rank-$s$
totally symmetric tensors $\vf_{\mu_1 \ldots \; \mu_s}$, that
generalize the metric tensor fluctuation $h_{\mu\nu}$ and possess
the gauge transformations
\be \delta \, \vf_{\mu_1 \ldots \; \mu_s} \, = \, \pr_{\mu_1} \,
\L_{\mu_2 \ldots \; \mu_s} \ + \ \ldots \ , \ee
with parameters $\L_{\mu_1 \ldots \; \mu_{s-1}}$ that are themselves
 totally symmetric tensors, of rank-$(s-1)$, or to rank-$n$ totally
symmetric spinor-tensors $\psi_{\mu_1 \ldots \; \mu_n}$, that
generalize the gravitino field $\psi_\mu$ of supergravity and
possess the gauge transformations
\be \delta \, \psi_{\mu_1 \ldots \; \mu_n} \, = \, \pr_{\mu_1} \,
\e_{\mu_2 \ldots \; \mu_n} \ + \ \ldots  \, , \ee
with parameters $\e_{\mu_1 \ldots \; \mu_{n-1}}$ that are themselves
totally symmetric spinor-tensors, of rank-$(n-1)$. For this class of
higher-spin fields, explicit statements can often be made
efficiently and concisely for arbitrary values of $s$ or $n$. It
should be borne in mind, however, that these types of fields do not
exhaust all available possibilities in more than four dimensions,
and indeed tensors of mixed symmetry are an important part of the
massive spectrum of String Theory. Previous results concerning the
free theory have been consistently generalized to cases of mixed
symmetry, albeit necessarily in a less explicit fashion, and hence,
for the sake of clarity, here we shall restrict ourselves to the
totally symmetric case, leaving for future work a detailed analysis
of the more involved cases of mixed symmetry. Abiding to common
practice, from now on we shall simply use the term ``spin'' for the
rank of bosonic tensors, or for the rank of Fermi fields augmented
by $1/2$.

The conventional formulation for free totally symmetric tensor gauge
fields $\vf_{\mu_1 \ldots \; \mu_s}$ was originally deduced by
Fronsdal \cite{fronsdal}, in the late seventies, from the massive
Singh-Hagen Lagrangians \cite{singhag}. The key feature of this
formulation is the need for a pair of constraints, one on the gauge
parameter, $\L_{\mu_1 \ldots \; \mu_{s-1}}$, whose \emph{trace}
$\L'_{\mu_3 \ldots \; \mu_{s-1}} \, \equiv \, \eta^{\mu_1\mu_2}\,
\L_{\mu_1 \ldots \; \mu_{s-1}}$ is required to vanish, and one on
the gauge field itself, whose \emph{double trace} $\vf^{\;
\prime\prime}_{\mu_5 \ldots \; \mu_s}\, \equiv \,
\eta^{\mu_1\mu_2}\, \eta^{\mu_3\mu_4}\, \vf_{\mu_1 \ldots \; \mu_s}$
is also required to vanish. In a similar fashion, the conventional
formulation for free totally symmetric spinor-tensors, due to Fang
and Fronsdal \cite{fang}, requires that both the
\emph{$\gamma$~-~trace} $\esl_{\mu_2 \ldots \; \mu_{n-1}}$ of the
gauge parameter and the \emph{triple $\gamma$~-~trace}\footnote{For
symmetric spinor-tensors two $\gamma$~-~traces are equivalent to a
trace.} of the spinor gauge field $\psisl^{\; \prime}_{\mu_4 \ldots
\; \mu_n}$ vanish. While these constraints result in a consistent
free dynamics, it is difficult to regard them as natural ingredients
of a complete formulation of higher-spin gauge fields.

There is a body of evidence that consistent higher-spin interactions
generally require that an infinite number of such fields be mutually
coupled. For instance, the gravitational coupling for a single
higher-spin field suffers from inconsistencies, the so-called
Aragone-Deser \cite{ades} problem, that can only be avoided in
special cases, and most notably in (A)dS backgrounds. Only in such
special circumstances can these fields be considered in isolation as
in flat space. On the other hand, interacting systems of higher-spin
fields are bound to be very complicated, and hence are not fully
under control to date, but possess the intriguing virtue of being of
intermediate complexity between ordinary low-spin Field Theory and
String Theory. Still, many important results are now available. Most
notably, in the presence of a cosmological term, the perturbative
definition of higher-spin interactions around (A)dS spaces can avoid
the difficulties long recognized for their naive flat-space
couplings \cite{fradkvas}. This crucial observation led Vasiliev to
formulate a consistent set of coupled non-linear equations for
totally symmetric tensors, first in four dimensions and, more
recently, in arbitrary dimensions as well. The Vasiliev equations
\cite{vas,vasnew} (see also \cite{sezsund} for relevant
contributions along these lines) represent the most encouraging
result on higher-spin gauge fields available to date, although they
are clearly non-Lagrangian and more work is required to arrive at an
off-shell formulation. They generalize both the frame formulation of
Einstein gravity and the Cartan integrable systems that have long
emerged from supergravity \cite{dfre}, extending them to allow for
non-polynomial scalar couplings.

The Vasiliev equations are based on an extension of the frame
formalism for gravity, and as a result their fields, forms valued in
representations of the tangent Lorentz group, can simply accommodate
Fronsdal's constraints. Still, other possibilities should be
explored at this stage, and in metric-like formulations the trace
constraints of \cite{fronsdal,fang} should naturally be absent. With
this motivation in mind, in \cite{fs1} we showed that it is possible
to extend the Fronsdal construction to allow for unconstrained gauge
fields and parameters. A nice outcome was the prompt emergence of
the geometry underlying the field equations, that become
\be \frac{1}{\Box^{p}} \ \prd \, {\cal R}^{[p]}{}_{; \; \alpha_1
\cdots \alpha_{2p+1}} \  =\  0 \ , \label{intro1} \ee
for odd spins $s=2p+1$, and
\be \frac{1}{\Box^{p-1}} \ {\cal R}^{[p]}{}_{; \; \alpha_1 \cdots
\alpha_{2p}} \ =\ 0 \ , \label{intro2} \ee
for even spins $s=2p$, where ${\cal R}$ denotes the linearized
higher-spin curvature introduced by de Wit and Freedman in
\cite{dewfr} and the bracketed suffix denotes that $p$~-~fold traces
are taken. Being \emph{non local} for spin $s>2$, both these
equations and the corresponding Lagrangians are not easy to use, but
their geometrical nature is nonetheless quite suggestive. Similar
constructions for higher-spin fermions were also presented in
\cite{fs1}, and the bosonic construction of \cite{fs1} was
generalized to mixed symmetry tensors in \cite{mixed}.

Local field equations for unconstrained fields can actually be
obtained without much effort. Confining our attention for simplicity
to bosonic fields, it possible to show \cite{fs2,st} that the
non-local geometric equations (\ref{intro1}) and (\ref{intro2}) are
equivalent to simple non-Lagrangian systems involving a new field, a
spin~-~$(s-3)$ compensator $\a_{\m_1 \ldots \; \mu_{s-3}}$, that
under gauge transformations transforms as
\be \delta \, \a_{\m_1 \ldots \; \mu_{s-3}} \, =\, \L^{\;
\prime}_{\m_1 \ldots \; \mu_{s-3}} \ . \ee
As we shall review in the next Section, the resulting local
compensator form of the higher-spin equations is equivalent to
eqs.~(\ref{intro1}) and (\ref{intro2}), and is actually suggested by
String Field Theory \cite{sft}. It can also be extended in a
relatively simple fashion to (A)dS backgrounds, and similar results
apply to its fermionic analog of \cite{fs2,st}.

The role of the unconstrained gauge symmetry in the interactions of
higher-spin gauge fields is less clear at the moment, but there are
clues that off-shell they will eventually make use of it. To wit,
while the four-dimensional Vasiliev construction of \cite{vas} was
based on a generalization of the two-component formalism for
gravity, and as a result is strictly tied to the Fronsdal form, this
is not necessarily the case for the more recent construction of
\cite{vasnew}. As stressed in \cite{sss}, in some respects the new
formulation of \cite{vasnew} can be regarded as a step forward in
the direction of an off-shell formulation, and indeed it allows to
drop the trace conditions. When this is done \cite{sss}, at the free
level one recovers very nicely the local compensator equations of
\cite{fs2,st}. At the interacting level, however, this choice
entails some subtleties that are spelled out in detail in
\cite{sss,bkiv}, and further work is required to settle the issue of
its consistency, although a forthcoming microscopic analysis of the
Vasiliev equations lends further, independent support to the role of
the unconstrained symmetry \cite{es}. Recently, additional evidence
for the potential role of the unconstrained symmetry for the
off-shell description of higher-spin gauge fields was also provided
in \cite{vas05}.

Complete off-shell formulations for free symmetric tensors were
already introduced some time ago by Pashnev and Tsulaia \cite{pt}.
Like their more recent fermionic counterparts of \cite{bp}, these
constructions rest on the BRST formalism and describe spin~-~$s$
symmetric tensors via additional fields whose number grows
proportionally to $s$. The resulting free systems are rather
complicated, but nonetheless in \cite{st} it was shown that a
judicious elimination of most of the additional fields reproduces
the geometric equations (\ref{intro1}) and (\ref{intro2}) in the
compensator form of \cite{fs2,st}. The present letter is devoted to
displaying far simpler constructions, minimal Lagrangians where the
trace constraints of the conventional Fronsdal formulation are
eliminated adding \emph{only two} fields, the compensator $\a$ and a
single spin~-~$(s-4)$ symmetric tensor $\beta$ playing the role of a
Lagrange multiplier. The next Section reviews briefly the results of
\cite{fs1,fs2,st}, while the two remaining Sections are devoted,
respectively, to the minimal bosonic Lagrangians and to their
fermionic counterparts.


\scs{Geometric and Local Compensator Equations}\label{sec:2}


As anticipated in the Introduction, in this paper we restrict our
attention to an important class of higher-spin fields, totally
symmetric (spinor-)tensors, a choice that has the virtue of allowing
a relatively handy discussion. The Fronsdal equations are
\be {\cal F} \ \equiv \ \Box \, \vf \ -\  \partial \, \partial \cdot
\vf \ +  \ \partial^2 \, \vf^{\, \prime} \, = \, 0 \ ,
\label{fronsdeq} \ee
where ${\cal F}$ will be often referred to as the Fronsdal operator.
For spin one and two, eqs.~(\ref{fronsdeq}) reduce to the Maxwell
equation for the vector potential $A_\mu$ and to the linearized
Einstein equation for the metric fluctuation $h_{\mu\nu}$, while
novelties begin to emerge for spin 3.

Let us pause briefly to explain our notation. In this letter, as in
\cite{fs1,fs2,st}, primes (or bracketed suffixes) denote traces,
while all indices carried by the symmetric tensors $\vf_{\mu_1 \dots
\mu_s}$ and $\L_{\mu_1 \dots \mu_{s-1}}$, by the metric tensor
$\eta_{\mu\nu}$ or by derivatives are left implicit. In order to
fully profit from this shorthand notation, where all terms are meant
to be totally symmetrized so that, for instance, $\pr \, \vf$ stands
for $\partial_{\mu_1} \vf_{\mu_2 \ldots \mu_{s+1}} \, + \,
\ldots\;$, one need only get accustomed to a few rules, that is
convenient to display again, correcting also a misprint in
\cite{fs2}:
\be
\begin{split}
  \left( \pr^{\, p} \, \vf  \right)^{\, \pe} \ &= \, \Box \,
  \pr^{\, p-2} \, \vf \ + \, 2 \, \pr^{\, p-1} \,  \prd \vf \ + \, \pr^{\, p} \,
\vf^{\, \pe} \,  , \\
 \partial^{\, p} \ \partial^{\, q} \ &= \ \binom{p+q}{p} \
\partial^{\, p+q} \ ,
\\
 \partial \cdot  \left( \partial^{\, p} \ \vf \right) \ & = \ \Box \
\partial^{\, p-1} \ \vf \ + \
\partial^{\, p} \ \partial \cdot \vf \ ,  \\
 \partial \cdot  \eta^{\, k} \ &= \ \partial \, \eta^{\, k-1} \ , \\
 \left( \eta^k \, \vf  \,  \right)^{\, \prime} \ &=  \, \left[ \, D
\, + \, 2\, (s+k-1) \,  \right]\, \eta^{\, k-1} \, \vf \ + \ \eta^k
\, \vf^{\, \prime} \ . \label{etak}
\end{split}
\ee
We shall work throughout in $D$ dimensions, with a mostly positive
space-time signature, and in this notation the Fronsdal Lagrangian
is simply
\be \label{fronlagr} {\cal L}_0 \, = \, \12 \ \vf \left(\ {\cal F}
\, - \, \frac{1}{2} \, \h \, {\cal F}^{\, \prime} \right) \ , \ee
where, as will be always the case in the following, evident
contractions between different fields in Lorentz invariant monomials
are left implicit. For instance, here $\vf$ is implicitly contracted
with the Einstein-like tensor ${\cal F} \, - \, \frac{1}{2} \, \h \,
{\cal F}^{\, \prime}$.

This formulation rests crucially on two constraints, that first
emerge for spin 3 and 4. The first concerns the gauge parameter,
$\L$, that enters the gauge transformation of $\vf$ in the
conventional fashion, as $\delta\,  \vf \, = \, \partial \, \L$, but
is to be {\it traceless} in order to guarantee the gauge invariance
of (\ref{fronsdeq}), since
\be \delta {\cal F} \, = \, 3 \ \pr^{\, 3} \L' \ . \ee
The second concerns the gauge field $\vf$ itself, that is to be {\it
doubly traceless}. This peculiar restriction originates from the
``anomalous'' Bianchi identity for the Fronsdal operator,
\be
\partial \cdot {\cal F} \, - \, \frac{1}{2} \, \partial \, {\cal F}^{\, \pe} \, = \, -
\ \frac{3}{2} \, \pr^{\, 3} \, \vf^{\, \prime\prime} \, ,
\label{bianchibose} \ee
since indeed, even with a traceless $\Lambda$, and up to partial
integrations that will be always left implicit in the following, the
Lagrangian (\ref{fronlagr}) \emph{is not} gauge invariant, but
varies into
\be \label{lfronvar} {\delta \cal L}_0 \, = \, - \ \frac{s}{2} \ \L
\left(\ \pr \cdot {\cal F} \ - \ \frac{1}{2} \ \pr \ {\cal
F}^{\;\prime} \right) \ . \ee

In \cite{fs1} we showed that, making use of the gauge field $\vf$
only, one can build a sequence of pseudo-differential analogs of
${\cal F}$,
\be {\cal F}^{(k+1)} \, = \, {\cal F}^{(k)} \ + \ \frac{1}{(k+1) (2
k+ 1)} \ \frac{\partial^{\;2}}{\Box} \, {{\cal F}^{(k)}}^{\;\prime}
\ - \ \frac{1}{k+1} \ \frac{\partial}{\Box} \
\partial \cdot  {\cal F}^{(k)} \ , \label{recursion} \ee
such that
\begin{equation}
\delta {\cal F}^{(k)} \, = \, \left( 2 k + 1 \right) \ \frac{
\partial^{\; 2 k + 1}} {\Box^{\; k-1}} \ \Lambda^{[k]} \ .
\end{equation}
For spin $s=2k-1$ or $s=2k$, ${\cal F}^{(k)} \, = \, 0$ is thus the
minimal fully gauge invariant modification of the Fronsdal equation.
Interestingly, this result is directly linked to the higher-spin
curvatures $\cR$ introduced by de Wit and Freedman \cite{dewfr}, and
indeed the fully gauge invariant equations can be written in the
more suggestive \emph{geometric} fashion \cite{fs2}
\be \frac{1}{\Box^{p}} \ \prd  {\cal R}^{[p]}{}_{; \; \alpha_1
\cdots \alpha_{2p+1}} \  =\  0 \ , \label{oddcurv} \ee
for odd spins $s=2p+1$, and
\be \frac{1}{\Box^{p-1}} \ {\cal R}^{[p]}{}_{; \; \alpha_1 \cdots
\alpha_{2p}} \ =\ 0 \ , \label{evencurv} \ee
for even spins $s=2p$.

In this formulation gauge invariance is manifest, and does not
require any constraints on the gauge parameter, but the equivalence
to the Fronsdal formulation entails a few subtleties, that are
spelled out in \cite{fs2}. In addition, the Bianchi identity is also
modified, so that
\be
\partial \cdot {\cal F}^{\,(k)} \, - \, \frac{1}{2k} \
\partial {{\cal F}^{\,(k)}}^{\, \prime} \, = \, - \,
\left( 1 + \frac{1}{2k}  \right) \, \frac{\partial^{\;
2k+1}}{\Box^{\; k-1}} \ \vf^{[k+1]}  \, , \label{bianchin} \ee
and this suffices to show that, for every given spin, there exists a
lowest value of $k$ such that the generalized Einstein tensors
\be {\cal G}^{(k)} \, = \, \sum_{p \leq k} \ \frac{(-1)^p}{2^p \
p{\;!} \ \binom{k}{p}} \ \eta^{\, p} \ {\cal F}^{(k)\, [p]}  \ee
are divergence-free, and hence the Lagrangians
\be {\cal L} \, = \, \frac{1}{2} \ \vf \ \cG^{(k)} \label{nonloclag}
\ee
are fully gauge invariant.

In \cite{fs1} we also showed that the geometric equations
(\ref{oddcurv}) and (\ref{evencurv}) can be always turned into the
form
\be {\cal F} \, = \, 3\, \pr^3 \, \cH \ , \ee
with $\cH$ a \emph{non-local construct} of ${\cal F}$, bound to
transform as $\delta \cH \, = \, \L'$ under a gauge transformation.
In other words, $\cH$ behaves as a \emph{compensator} for the trace
of the gauge parameter.

In \cite{fs2,st}, drawing also from String Field Theory \cite{sft},
we explored the implications of allowing in the theory an
\emph{independent field} $\alpha$, denoted as the ``compensator''
and such that $\delta \alpha \, = \, \L'$. The key result of this
analysis was that with $\vf$ and $\a$ one can arrive at the two
local field equations
\bea
{\cal F} &=& 3\, \pr^{\, 3} \a \, , \label{compeq1} \\
\vf^{\, \pe \pe} &=& 4 \, \pr \cdot \a \ + \ \pr \, \a^{\, \pe} \, ,
\label{compeq2} \eea
that can also be obtained truncating the bosonic triplet of
\cite{oldtriplet,pt}, are nicely consistent with the Bianchi
identity (\ref{bianchibose}), but \emph{are not} Lagrangian. As
shown in \cite{fs2} (see also \cite{tensorial}), from these one can
readily recover the non-local geometric equations (\ref{oddcurv})
and (\ref{evencurv}) building a sequence of equations for the
non-local extensions ${\cal F}^{(k)}$ of (\ref{recursion}), since
eq.~(\ref{compeq1}) implies that
\be {\cal F}^{(k)} \, = \, (2 \, k \ + \ 1) \ \frac{\pr^{\, 2
k+1}}{\Box^{k}} \, \alpha^{[k]} \ , \ee
and finally, after the minimal number of iterations needed to
produce a trace of $\a$ not allowed for a given spin $s$,
eqs.~(\ref{oddcurv}) and (\ref{evencurv}).

In a similar fashion starting from the fermionic Fang-Fronsdal
operator
\be {\cS} \, = \, i\, \left( \dsll \, \psi \, - \, \pr \, \psisl
\right) \, , \ee
one can define a sequence of non-local extensions of $\cS$, directly
linked to the bosonic ones according to
\be {\cal S}^{(k)}_{n+1/2} \ - \ \frac{1}{2k} \, \frac{\pr}{\Box}\,
{\not{\!\pr}} \, \ssl_{n+1/2}^{(k)} \, = \, i \
\frac{\not{\!\pr}}{\Box} \, {\cal F}^{(k)}_n(\psi) \, , \ee
with ${\cal F}^{(k)}_n(\psi)$ a non-local extension of the Fronsdal
operator for the spinor-tensor $\psi$, and arrive eventually at
non-local geometric equations. Even in this case, non-Lagrangian
equations for spin~-~$(n+1/2)$ spinor-tensors $\psi_{\mu_1 \ldots
\mu_n\;}$ involving a single spin~-~$(n-3/2)$ compensator
$\xi_{\mu_1 \ldots \mu_{n-2}\;}$ were obtained in \cite{fs2,st}. In
flat space they read
\bea
{\cal S} &=& - \ 2 \, i \, \partial^2 \, \xi \ , \nonumber \\
\psisl^{\ \pe} &=& 2 \, \partial \cdot \xi \ + \ \partial \, \xi^{\,
\pe} \ + \ {\not{\!\pr}} \xisl \ , \label{compfermiflat} \ \eea
and can be obtained truncating the fermionic triplet introduced in
\cite{fs2}.


\scs{Local Lagrangians for unconstrained bosons}\label{sec:2.2}


In the previous Section we have reviewed the salient features of the
non-local geometric equations of \cite{fs1} and their relation with
the non-Lagrangian equations of \cite{fs2,st}. As shown in
\cite{st}, the latter also follow, albeit in a somewhat indirect
fashion, from the BRST construction of Pashnev and Tsulaia
\cite{pt}.

In this Section we would like to present a simple alternative: local
Lagrangians for unconstrained spin~-~$s$ tensor fields $\varphi$
that involve at most {\it two} additional fields, the spin~-~$(s-3)$
compensator $\alpha$ of \cite{fs2,st}, and an additional
spin~-~$(s-4)$ field $\beta$ that acts as a Lagrange multiplier for
the relation between the double trace $\varphi^{\, \prime\prime}$
and the compensator $\alpha$ in eq.~(\ref{compeq2}). These
``minimal'' Lagrangians are closely related to the geometric
equations (\ref{oddcurv}) and (\ref{evencurv}) of \cite{fs1} via the
compensator system (\ref{compeq1}) and (\ref{compeq2}), and for spin
$s=3$ reduce to the result already presented in \cite{fs1}.

A minimal gauge invariant Lagrangian for spin~-~$s$ symmetric
tensors $\vf$ can be nicely determined resorting to the familiar
Noether procedure, that allows one to deal with this problem in a
systematic fashion and has the additional virtue of clarifying the
origin of the difficulty met in a naive approach beyond the spin~-~3
case. Let us therefore begin by considering the Fronsdal expression
(\ref{fronlagr}), now written for a field $\varphi$ \emph{not
subject} to any trace constraints, and let us vary it without
enforcing any constraints on the gauge parameter $\L$. The resulting
complete variation,
\be
\begin{split}
\d {\cal{L}}_0 \, =& - 3\, \4s \, \vf^{\, \prime \prime} \, \prd
\prd \prd \L \, -9 \, \4s \, \prd \prd \vf^{\, \prime}\, \prd \L^{\,
\prime} \,
+\, \fr{15}{2} \5s \prd \prd \prd \vf^{\, \prime} \, \L^{\, \prime \prime} \\
&+ \, \L^{\, \pe} \, \3s \, \left\{ \fr{3}{4} \, \prd {\cal{F}}^{\,
\pe} - \, \fr{3}{2} \, \prd \prd \prd \vf \, +\, \fr{9}{4} \Box \,
\prd \vf^{\, \pe} \, \right\} \, ,
\end{split}
\ee
comprises a number of terms depending on $\Lambda'$, that can be
canceled adding
\be
\begin{split}
 \cL_1 \, = & \, - \, \a \, \3s \, \left\{ \fr{3}{4} \, \prd {\cal{F}}^{\, \pe} \, - \, \fr{3}{2}
 \, \prd \prd \prd \vf \,
 +\, \fr{9}{4} \, \Box \, \prd \vf^{\, \pe} \, \right\} \\
                & + \, 9\, \4s \, \prd \a \, \prd \prd \vf^{\, \pe} \, - \, \fr{15}{2}
 \5s \a^{\pe} \, \prd \prd \prd \vf^{\, \pe} \, ,
\end{split}
\ee
that depends linearly on the compensator $\alpha$.

Additional terms depending on $\L'$ now present themselves in the
resulting variation of ${\cal{L}}_0 + {\cal{L}}_1$, but can be
eliminated adding
\be
\begin{split}
{\cal{L}}_2 \, = \, & \, \fr{9}{4} \, \3s \a \, \Box^2 \, \a -\, 27
\, \4s \, \prd \a \, \Box
\, \prd \a \, + \, 45 \5s (\prd \prd \a)^2 \\
& +\, \fr{45}{2} \5s \prd \prd \a \, \Box \, \a^{\, \pe} \, - \, 45
\, \6s \, \prd \prd \prd \, \a \, \prd \a ^{\, \pe} \, ,
\end{split}
\ee
that is quadratic in $\a$, so that the final remainder is
\be \d \left\{ {\cal{L}}_0  \, + \, {\cal{L}}_1 \, +\, {\cal{L}}_{2}
\right\} \, = \, - \ 3 \, \4s \, \left\{\vf^{\, \pe \pe} -\, 4 \prd
\a \ - \pr \a^{\, \pe} \right\} \, \prd \prd \prd \L \, . \ee
These terms vanish for $s<4$, and are proportional to the gauge
invariant expression given by eq.~(\ref{compeq2}). A fully gauge
invariant unconstrained Lagrangian is thus finally obtained
introducing, from spin $s=4$ onwards, the single additional term
\be {\cal{L}}_{3} \, = \, 3\, \4s \, \b \, \left( \vf^{\, \pe \pe}
\,-\, 4 \prd \a \, - \, \pr \a^{\, \pe} \right)\, , \ee
where the spin~-~$(s-4)$ \emph{Lagrange multiplier} $\b$ transforms
as $\d \b \, = \, \prd \prd \prd \, \L$.

Summarizing, the complete Lagrangians for unconstrained spin~-~$s$
totally symmetric tensors are
\be
\begin{split}
{\cL} = & \, \hspace{.1cm} \12\, \vf \, \left(\ {\cal F} \, - \, \12
\, \h \, {\cal F}^{\, \prime}\right)\, - \, {s \choose 3}\, \a \,
\left\{ \fr{3}{4} \, \prd {\cal F}^{\, \prime}\, -\, \fr{3}{2}\,
\prd
\prd \prd \vf \, +\, \fr{9}{4} \, \Box \, \prd \vf^{\, \prime} \right\} \\
&+\, 9\, {s \choose 4}\, \prd \a \, \prd \prd \vf^{\, \prime}\, - \,
\fr{15}{2} \, {s \choose 5} \, \a^{\, \prime}
\prd \prd \prd\vf^{\, \prime} \, + \, \fr{9}{4}\, \3s\, \a \, \Box^2 \, \a \\
&-\, 27\, \4s\, \prd \a\, \Box \, \prd \a \, + \, 45\, \5s \, (\prd
\prd \a)^2
+\, \fr{45}{2} \5s \, \prd \prd \a \, \Box \, \a^{\, \prime} \\
& - \, 45\, \6s \, \prd \prd \prd \a \, \prd \a^{\, \prime} +3\, \4s
\, \b \, \left(\vf^{\, \prime \prime} \, - \, 4\, \prd \a \, - \,
\pr \, \a^{\, \prime}\right) \ ,   \label{lagrange}
\end{split}
\ee
and are invariant under the gauge transformations
\be
\begin{split}
& \d \, \vf\, = \, \pr \, \L \, , \\
& \d \, \a \, = \, \L^{\, \prime} \, , \\
& \d \, \b \, = \prd \prd \prd \L \, .
\end{split}
\ee

We can now move on to clarify the connection with the non-Lagrangian
system of eqs.~(\ref{compeq1}) and (\ref{compeq2}). The starting
point are the field equations determined by (\ref{lagrange}),
\begin{align}
\vf  &:  \hspace{.5cm}    {\cal F} \, - \, 3 \, \pr^{\, 3} \, \a \,
- \, \12 \,  \h \, ({\cF}^{\, \pe} \, - \, \12 \, \pr^2 \, \vf^{\,
\prime
                                     \prime}\, - \, 3\, \Box\, \pr \, \a   \,-\, 4\,
                                     \pr^{\, 2}\, \prd \a -\, \fr{3}{2}\, \pr^{\, 3} \, \a^{\, \pe})  \nonumber \\
     &   \hspace{3cm}   + \,\h^2  \, (\b\, + \12 \, \pr\, \prd \prd \a \, +
     \, \Box\, \prd \a\, - \, \12 \, \prd \prd \vf^{\, \pe} )\, = \, 0\, , \label{phi} \\
     &   \nonumber \\
\b  &:  \hspace{.5cm}     \vf^{\, \pe \pe} \, - \, 4\, \prd \a \, - \, \pr \, \a^{\, \pe} \, = 0 \, , \label{beta} \\
     &   \nonumber \\
\a  &: \hspace{.5cm}      6\, \Box^2\, \a \,+\, 18\, \Box \, \pr\,
\prd \a \, +\, 12\, \pr^2 \, \prd \prd \a \, + \, 3\, \Box\, \pr^2
\,
                                      \a^{\, \pe}
\,+\, 3\, \pr^3\, \prd \a^{\, \pe} \nonumber \\
&  \hspace{.6cm}           -\, 3 \, \pr \, \prd \prd \vf^{\, \pe} \,
- \, \prd {\cal F}^{\, \pe} \,+\, 2\, \prd \prd \prd \vf \, - \,
3\, \Box \, \prd \vf^{\, \pe} \, + \, 4\, \pr \, \b  \nonumber \\
&  \hspace{.6cm}           +\, \h \, ( 3\, \Box\, \prd \prd \a \,+\,
\pr\,  \prd \prd \prd \a \,-\, \prd \prd \prd \vf ^{\, \pe} \,
 +\, 2\, \prd \b) \,=\,0\, ,   \label{alfa}
 \end{align}
and the issue is to show their equivalence to eq.~(\ref{compeq1}). A
general argument to this effect can be built as follows.

Let us begin by noticing that, when $\b$ is on-shell, {\it i.e.}
when eq.~(\ref{beta}) is enforced, the $\vf$ equation becomes of the
form
\be \label{abc} \cA\, - \, \12 \, \h\, \cA^{\, \pe} \, + \, \h^2 \,
\cC \, = \, 0 \, , \ee
where
\be \label{ABC}
\begin{split}
\cA \, &= \, {\cal F} \, - \, 3 \, \pr^{\,3} \, \a\, , \\
\cC \, &= \, \b\, + \, \12 \, \pr\, \prd \prd \a \, + \, \Box\, \prd
\a\, - \,\12 \, \prd \, \prd \, \vf^{\, \pe} \, ,
\end{split}
\ee
and that, under the same condition, the double trace of $\cA$
vanishes \emph{identically}. One can then take successive traces of
(\ref{abc}): whereas the first relates $\cA^{\, \pe}$ to $\cC$ and
$\cC^{\, \pe}$, the higher ones yield relations of the form
\be \label{eliminate} \left( \h^2 \, \cC \right)^{[k]}  \, = \, \h^2
\,\cC^{[k]}\,+ \, \sum_{i=1}^k\,\r_{2i+1} \, \h\, \cC^{[k-1]} \, +\,
\sum_{i\,\leq\, j =  2}^k \r_{2i-1} \, \r_{2j}\, \cC^{[k-2]} \, =\,
0\, . \ee
It should be appreciated that the $\cC^{[i]}$ are all independent
and do not vanish identically. As a result, these never reduce to
trivial identities, since the coefficients $\r_k \equiv
D\,+\,2\,(s\, -\, k)$ are positive. Therefore, denoting with $p$ the
integer part of $\fr{s - 4}{2}$ and taking $p+2$ traces of
(\ref{abc}) gives
\be \sum_{i\,\leq\, j = 2}^{p+2}\, \r_{2i-1} \, \r_{2j}\, \cC^{[p]}
\,  =\, 0\, , \ee
and hence, finally, $\cC^{[p]}\, = \, 0$. Making use of this
relation in the $[p+1]$-th trace, and working backwards, one can
convince oneself that \emph{on shell all traces of $\cC$, including
$\cC$ itself, vanish}. In the first trace of (\ref{abc}), this
result gives $\cA^{\, \pe} \, = \, 0 \,$, and then finally
eq.~(\ref{abc}) turns into the desired form
(\ref{compeq1})\footnote{$s=2$ and $D=2$ is a well-known exception,
since in that case the trace of $\cA \, -\, \frac{1}{2}\, \eta\,
\cA^{\, \prime}$ vanishes identically, giving no indications on
$\cA^{\, \prime}$.}.

 As we have seen, the field equation for the Lagrange
multiplier $\b$ is the condition that the double trace of the
dynamical field be pure gauge, and plays a crucial role in linking
these Lagrangian equations to the geometrical ones. On the other
hand, ({\ref{alfa}) has not played any role so far, in particular in
the relation between the local Lagrangian (\ref{lagrange}) and the
higher-spin geometry. There is a reason for this: (\ref{alfa}) is a
\emph{consequence} of the field equations for $\vf$ and $\b$.
Indeed, taking the divergence of eq.~(\ref{phi}) and using
eq.~(\ref{beta}) in the result, one arrives at an expression
proportional to (\ref{alfa}). More precisely, indicating with
$\cG_{\vf, \b}(\a)$ the field operator for the compensator $\alpha$,
one can see that
\be
 \prd\, \{\cA\, - \, \12 \, \h\, \cA^{\, \pe} \, + \, \h^2 \, \cC\} \, = \, \fr{\h}{4} \, \cG_{\vf, \b}(\a) \, .
 \ee
It is then clear that if $\vf$ and $\b$ satisfy their field
equations, $\cG_{\vf, \b}(\a)$ is forced to vanish. That is to say,
$\a$ is forced to satisfy its field equation as well.

Actually, the role of the field equation for the compensator $\a$
can be better appreciated if the dynamical field $\vf$ is coupled to
an external source $\cJ$. In the Fronsdal case, the Lagrangian
equation is
\be {\cal F} \ - \ \frac{1}{2} \ \h \ {\cal F}^{\, \pe}  \, = \, \cJ
\, , \ee
and its divergence gives
\be -\, \12 \, \h \, \prd \cF^{\, \pe} \, = \, \prd \cJ \, . \ee
Hence, while in the Maxwell and Einstein cases $\prd \cF^{\, \pe}$
vanishes identically, so that the sources must be \emph{divergence
free}, in the conventional formulation for spin 3 or higher only the
\emph{traceless part of the divergence} is forced to vanish. In
\cite{fronsdal} it was shown that even this weaker condition
suffices to ensure that only physical polarizations contribute to
the exchange of quanta between sources. On the other hand, if an
external source is introduced in our Lagrangian (\ref{lagrange}) via
the standard coupling $\vf \, \cdot \, \cJ$, the field equations
become
\begin{align}
&\cA\, - \, \12 \, \h\, \cB \, + \, \h^2 \, \cC \, = \cJ \, , \label{abc1} \\
&\cG_{\vf, \b}(\a) \, = \, 0 \, , \label{abc2} \\ &\vf^{\, \pe
\pe}\, - \, 4\,  \prd \a \, -\,  \pr\, \a^{\, \pe}\, = \, 0\, ,
\label{abc3}
\end{align}
where $\cB$ is defined by comparing with eq.~(\ref{phi}). Combining
the divergence of (\ref{abc1}) with (\ref{abc3}) then yields
\be \fr{\h}{4} \, \cG_{\vf, \b}(\a) \, = \, \prd \cJ \, , \ee
a result apparently similar to Fronsdal's \cite{fronsdal}. However,
now the full Lagrangian system implies that, when the compensator
$\a$ satisfies its field equation, the coupling can only be
consistent if the current ${\cal J}$ is \emph{divergence free}.
Incidentally, this is just the expected Noether constraint for a
source related to the gauge symmetry of the theory described by
(\ref{lagrange}).


\scs{Local Lagrangians for unconstrained fermions}


One can repeat the previous steps almost verbatim for fermion
fields. Here the starting point is provided by the Fang-Fronsdal
equation for a symmetric spin~-~$(n+1/2)$ spinor-tensor $\psi$,
\be \cS \ \equiv \ i\, \left( \dsll \psi \ - \ \pr \psisl \right) \
= \ 0 \ , \ee
that is invariant under the gauge transformation $\delta \, \psi \,
= \, \pr \, \e$ only if the gauge parameter is
\emph{$\gamma$~-~traceless}, since
\be \delta \ \cS \, = \, - \, 2\, i\, \pr^{\, 2} \! \esl \, . \ee
In a similar fashion, the corresponding Lagrangian
\be {\cal L} \, = \, \frac{1}{2} \ \bar{\psi} \left( \cS \, - \,
\frac{1}{2} \, \gamma \, \ssl \, - \, \frac{1}{2} \, \eta \, \cS^{\,
\pe} \right) \ + \ h.c. \label{fflagr} \ee
is gauge invariant only if the gauge field $\psi$ is \emph{triply
$\gamma$~-~traceless}, on account of the ``anomalous'' Bianchi
identity
\be \prd \cS\, - \, \frac{1}{2} \, \partial \ \cS^{\, \pe } \, - \,
\frac{1}{2} \, {\not {\! \pr}} \ssl \,
 = \, i\, \pr^{\; 2} \psisl^{\, \pe} \ .  \label{bianchifermi}
\ee

In this case one begins by considering the Lagrangian
(\ref{fflagr}), written however for an \emph{unconstrained} Fermi
field $\psi$ of spin $n+1/2$, and varied with an
\emph{unconstrained} gauge parameter $\e$. The resulting variation
is then
\be
\begin{split}
 \delta {\cal L}_0 =& - \ \frac{3 i}{2}\ {n \choose 3} \ \pr \cdot \pr \cdot \bar{\e} \ \psisl^{\ \pe} \ + \
 \frac{3i}{4} \ {n \choose 3} \ \bar{\e}^{\, \pe} \ \dsll \ \pr \cdot \psi^{\, \pe} \ - \ 3 i \ {n \choose 3} \ \bar{\e}^{\, \pe} \
 \prd \prd \psisl  \\
 & - \ \frac{3 i}{4} \ {n \choose 3} \ \bar{\e}^{\, \pe} \ \Box \ \psisl^{\ \pe} \ + \ 3i \ {n \choose 4} \
 \prd \bar{\e}^{\, \pe} \ \pr \cdot \psisl^{\ \pe} \ + \ 2 i \ {n \choose 2}\ \bar{\esl} \ \pr \cdot \pr \cdot \psi \\
 & - \ 2 i \ {n \choose 2} \ \bar{\esl} \ \dsll \ \pr \cdot \psisl \ - \ i \ {n \choose 2} \
\bar{\esl} \, \Box \, \psi^{\, \pe} \, + \, \frac{9 i}{2} \ {n
\choose 3} \ \pr \cdot \bar{\esl} \ \pr
\cdot \psi^{\, \pe} \\
& - \ 3 i \ {n \choose 4} \ \bar{\esl}^{\ \pe} \ \pr \cdot \pr \cdot
\psi^{\, \pe} \ + \ h.c. \, .
\end{split}
\ee
In complete analogy with the bosonic case, all terms involving the
$\gamma$~-~trace $\not \! \epsilon\, $ of the gauge parameter can be
canceled by additional terms linear in the compensator field $\xi$,
that are collected in
\be
\begin{split}
{\cal L}_1 =& - \, \frac{3 i}{4} \, {n \choose 3} \,\bar{\xisl} \
\dsll \, \pr \cdot \psi^{\, \pe} \ + \, 3 i \, {n \choose 3} \
\bar{\xisl} \, \prd \prd
\psisl \ + \ \frac{3 i}{4} \, {n \choose 3} \, \bar{\xisl} \, \Box \, \psisl^{\ \pe}  \\
                 & - \, 3i \, {n \choose 4} \, \pr \cdot \bar{\xisl} \, \pr \cdot \psisl^{\ \pe} \, - \, 2 i \
                 {n \choose 2}\,\bar{\xi} \, \prd \pr \cdot \psi  \, + \, 2 i \ {n \choose 2} \
                 \bar{\xi} \, \dsll \, \pr \cdot \psisl \\
                 & + \, i \, {n \choose 2} \, \bar{\xi} \, \Box \, \psi^{\, \pe} \, - \, \frac{9 i}{2} \, {n \choose 3} \
                  \pr \cdot \bar{\xi} \, \pr \cdot \psi^{\, \pe}\, + \, 3 i \,{n \choose 4} \, \bar{\xi}^{\, \pe} \, \prd
                  \prd \psi^{\, \pe} \, + \, h.c. \, .
\end{split}
\ee
Additional terms depending on $ \not{\!  \epsilon}\,$ generated by
the variation of ${\cal L}_1$ can then be eliminated adding
\be
\begin{split}
{\cal L}_2 \ =& - \, \frac{15 \, i }{2} \, {n \choose 3} \,
\bar{\xisl} \, \Box \, \prd \xi \, - \, i \, {n \choose 2} \,
\bar{\xi} \, \Box  \, \dsll \, \xi \, + \, 3 \, i \, {{n} \choose
{3}} \
 \prd \bar{\xi} \, \dsll \, \prd \xi \\
                    & + \ 18 \, i \, {n \choose 4} \, \prd \bar{\xisl} \, \prd \prd \xi \, + \, 6 \, i \, {n \choose 4} \
                    \prd \bar{\xisl} \,\Box \, \xi^{\, \pe}  \\
                    & - \, 15 \, i \, {n \choose 5} \, \prd \prd \bar{\xisl} \ \prd \xi^{\, \pe} \, + \, h.c. \, ,
\end{split}
\ee
and the total variation is finally
\be \label{finvar} \delta \left\{ {\cal L}_0 \ + \, {\cal L}_1 \, +
\, {\cal L}_2 \right\} \, = \, - \, \frac{3 \, i }{2} \, {n \choose
3} \ \pr \cdot \pr \cdot \bar{\e} \, ( \psisl^{\ \pe} \, - \, 2\,
\pr \cdot \xi \, - \, \dsll \xisl \, - \, \pr \xi^{\, \pe} ) \, + \,
h.c. \, . \ee

This residual contribution is proportional to the constraint
relating $\psisl^{\, \prime}$ to the compensator in
eqs.~(\ref{compfermiflat}). One can finally introduce a Lagrange
multiplier field $\lambda$, a spinor-tensor of spin $n \, - \, 5/2$
such that $\delta \, \lambda \, = \, \prd \prd \e$, to compensate
(\ref{finvar}) by the additional term
\be {\cal L}_3 \, = \,  \frac{3 \, i }{2} \, {n \choose {3}} \,
\bar{\lambda} \, \left( \psisl^{\ \pe} \, - \, 2\, \pr \cdot \xi \,
- \, \dsll \xisl \, - \, \pr \xi^{\, \pe}  \right) \, + \, h.c. \, .
\ee

Summarizing, the complete Lagrangian for an \emph{unconstrained}
spin~-~$(n+1/2)$ spinor-tensor is
\be \label{lagrferm}
\begin{split}
{\cal L}\, =\, & \hspace{2mm} \frac{1}{2} \, \bar{\psi} \left( \cS \, - \, \frac{1}{2} \, \gamma \, \ssl \, - \, \frac{1}{2} \, \eta \,  \cS^{\, \pe}  \right) \\
                   & - \, \frac{3 i}{4} \, {n \choose 3} \, \bar{\xisl} \, \dsll \, \pr \cdot \psi^{\, \pe} \, + \, 3 i \ {n\choose 3}
                    \ \bar{\xisl} \, \pr \cdot \pr \cdot \psisl \ + \, \frac{3 i}{4} \
                   {n \choose 3} \, \bar{\xisl} \, \Box \, \psisl^{\ \pe} \\
                   & - \, 3i \, {n \choose 4} \, \prd \bar{\xisl} \, \pr \cdot \psisl^{\ \pe} \, - \, 2 i \, {n \choose 2}\ \bar{\xi} \
                   \prd  \pr \cdot \psi \, + \ 2 i \, {n \choose 2} \, \bar{\xi} \, \dsll \, \pr \cdot \psisl \\
                   & + \, i \, {n \choose 2} \, \bar{\xi} \, \Box \, \psi^{\, \pe} \, - \, \frac{9 i}{2} \, {n \choose 3} \
                    \prd \bar{\xi} \, \pr \cdot \psi^{\, \pe} \, + \, 3 i \, {n \choose 4} \, \bar{\xi}^{\, \pe} \, \pr
                     \cdot \prd \psi^{\, \pe} \\
                   & - \, \frac{15 i }{2} \, {n \choose 3} \, \bar{\xisl} \, \Box \, \prd \xi \, - \, i \, {n \choose 2} \, \bar{\xi}
                  \, \Box  \, \dsll \, \xi \, + \, 3 i \, {n \choose 3} \, \prd \bar{\xi} \, \dsll \, \prd \xi  \\
                   & + \, 18 i \, {n \choose 4} \, \prd \bar{\xisl} \, \prd \prd \xi \, + \, 6 i \, {n \choose 4} \, \prd \bar{\xisl}
                   \, \Box \, \xi^{\, \pe} \, - \, 15 i \, {n \choose 5} \, \prd \prd \bar{\xisl} \, \prd \xi^{\, \pe} \\
                   & + \, \frac{3 i }{2} \, {n \choose 3} \, \bar{\lambda} \, \left( \psisl^{\ \pe} \, - \, 2\, \pr \cdot \xi \, - \, \dsll
                   \xisl \, - \, \pr \xi^{\, \pe}  \right) \ + \ h.c. \, ,
\end{split}
\ee
and is invariant under the gauge transformations
\be
\begin{split}
& \d \, \psi\, = \, \pr \, \e \ ,  \\
& \d \, \xi \, = \, \esl \ ,  \\
& \d \, \lambda \, = \prd \prd \e \ .
\end{split}
\ee

The corresponding field equations are:
\begin{align}
\bar{\psi} \,        &:  \hspace{.3cm}  - \, i \, \cS + 2 \, \pr^2
\xi \, - \, \frac{1}{2} \, \eta \,( -\, i\, \cS^{\, \pe} \, + \,
\frac{1}{2} \, \pr \psisl^{\, \pe} \, - \,
                            \frac{1}{2} \, \pr \, \dsll \, \xisl \, + \, 2 \, \Box \, \xi \, + \, 3\, \pr \prd \, \xi \, + \, \pr^2 \,
                            \xi^{\, \pe} )   \nonumber  \\
                          &  \hspace{.5cm}-\frac{1}{2} \, \gamma \, ( -\, i \, \ssl \, + \, 2 \, \pr^2 \xisl \, + \, 2\, \pr \dsll \, \xi ) \, +
                      \, \frac{1}{4} \, \gamma \, \eta \,
                      (\prd \psi^{\, \pe} \, - \, \Box \, \xisl \, - \, \pr \prd \xisl \, - \, 2\, \lambda )\, = \, 0\ , \label{psi} \\
                          & \nonumber \\
 \bar{\lambda} \, & : \hspace{.5cm}   \psisl^{\ \pe} \, - \, 2\, \prd \xi \, - \, \dsll \, \xisl\, - \, \pr \xi^{\, \pe} \, = \, 0 \ ,\label{lambda}  \\
& \nonumber \\
\bar{\xi} \,          &:  \hspace{.3cm} \Box\,  \psi^{\, \pe} \, +
\, 2\, \dsll \, \prd \psisl \, - \, 2\, \prd \prd \psi \, +\,
\frac{3}{2} \, \pr \prd
                     \psi^{\, \pe} \, -
\, 2 \, \Box \, \dsll \xi \, - \, \frac{5}{2} \, \Box\,
                     \pr \xisl \, - \, 2\, \dsll \, \pr \prd \xi \,   \nonumber \\
                 &  \hspace{.5cm} - \, 3\, \pr^2 \prd \xisl \, - \, \pr \lambda\,  +\, \eta \, (\frac{1}{2}\, \prd \prd \psi^{\, \pe} \,
                     - \, \Box \, \prd \xisl \, - \, \12 \, \pr \prd \prd \xisl \, - \, \prd \lambda) \nonumber \\
                &  \hspace{.5cm}  +\, \gamma \, ( - \, \frac{1}{4} \, \dsll \prd
                     \psi^{\, \pe}\, + \, \prd \prd \psisl \, + \, \frac{1}{4} \, \Box \,
                     \psisl^{\ \prime}
                     \, + \, \frac{1}{4} \, \pr \prd \psisl^{\; \prime}  \nonumber \\ 
                &  \hspace{.5cm} - \, \frac{5}{2} \, \Box \, \prd \xi \, - \,    
                   \frac{3}{2} \, \pr \prd \prd \xi \, - \, \frac{1}{2} \, \Box \, \pr \, \xi^{\, \pe}
                   \, - \, \frac{1}{2}\, \pr^2 \prd \xi^{\, \pe} \, - \, \frac{1}{2} \, \dsll \l ) \, = \, 0 \ . \label{xi}
\end{align}
As in the bosonic case, we can now relate them to the simple
non-Lagrangian system (\ref{compfermiflat}). The basic observation
is to recognize that, when (\ref{lambda}) is satisfied, the equation
for $\psi$ takes the form
\be \label{abferm} \cW \, -\, \12 \, \gamma \, \cWsl \, - \, \12 \,
\h \, \cW^{\, \pe} \, + \, \fr{i}{4} \, \h \, \gamma \, \cZ \, = \,
0 \, , \ee
where
\be
\begin{split}
\cW & = \cS \, + \, 2\, i \, \pr^2 \, \xi \, , \\
\cZ & = \prd \psi^{\, \pe} \, - \, \Box \, \xisl \, - \, \pr \, \prd
\xisl \, - \, 2\, \lambda \, .
\end {split}
\ee
Moreover, under the same conditions the triple $\gamma$~-~trace of
$\cW$ vanishes. It is then possible to rephrase the iterative
argument presented for bosonic fields: if $p$ is the integer part of
$\fr{s-3}{2}$, where $s\geq 3$, taking $(p+1)$ successive traces of
(\ref{abferm}) one arrives at the condition that the highest
$\gamma$~-~trace vanish:
\be \gamma \cdot \left(\fr{1}{4} \, \h \, \gamma \, \cZ
\right)^{[p+1]} \, = \, 0\, . \label{maxgamma} \ee
Inserting (\ref{maxgamma}) in the lower $\gamma$~-~traces of
(\ref{abferm}), one can recursively show that \emph{all}
$\gamma$~-~traces of $\cZ$ vanish as well. Consequently, the first
trace and the $\gamma$~-~trace of (\ref{abferm}) imply that $\cW^{\,
\pe}\, = \, 0$ and $\not \! \! \! \cW \, = \, 0$, and in conclusion
the Lagrangian (\ref{lagrferm}) leads indeed to the local
compensator equations (\ref{compfermiflat}). In complete analogy
with the bosonic case, making use of (\ref{lambda}) one can show
that the field equation (\ref{xi}) for the compensator $\xi$ is
proportional to the divergence of (\ref{psi}). \vskip 12pt

\section*{Acknowledgments}


We are grateful to the CPhT of the Ecole Polytechnique, to the
LPT-Orsay and to the CERN Theory Division for the kind hospitality
extended to us, and to C.~Iazeolla, J.~Mourad and M.~Porrati for
stimulating discussions. The visits of A.S. were partly supported by
a ``poste rouge'' CNRS, by the CNRS PICS n.~3059 and by the CERN TH
unit, while the visits of D.F. were partly supported by the CNRS
PICS n. 3059 and by the APC group of U. Paris VII. The present work
was also supported in part by INFN, by the MIUR-COFIN contract
2003-023852, by the EU contracts MRTN-CT-2004-503369 and
MRTN-CT-2004-512194, by the INTAS contract 03-51-6346, and by the
NATO grant PST.CLG.978785.

\vskip 12pt


\end{document}